\documentclass[fleqn,usenatbib]{mnras}

\usepackage[T1]{fontenc}

\DeclareRobustCommand{\VAN}[3]{#2}
\let\VANthebibliography\thebibliography
\def\thebibliography{\DeclareRobustCommand{\VAN}[3]{##3}\VANthebibliography}

\usepackage{graphicx}	
\usepackage{amsmath}	
\usepackage{amssymb}	
\usepackage{gensymb}
\usepackage{newtxtext,newtxmath}

\newcommand{\Mjup}{\mbox{M$_{\mathrm{jup}}$}}
\newcommand{\Msun}{\mbox{M$_{\odot}$}}
\newcommand{\Rsun}{\mbox{R$_{\odot}$}}

\newcommand{\lppr}{\stackrel{<}{\scriptstyle \sim}}
\newcommand{\lappr}{\raisebox{-0.4ex}{$\lppr$}}

\title[Origin of a close brown dwarf around a WD]{Common envelope evolution and triple dynamics as potential pathways to form the inner white dwarf + brown dwarf binary of the triple star system Gaia\,0007$-$1605}

\author[Lagos et al.]{
Felipe Lagos$^{1}$\thanks{Felipe.Lagos-Vilches@warwick.ac.uk},
Monica Zorotovic$^{2}$,
Matthias R. Schreiber$^{3,4}$
B. T. G\"ansicke,$^{1, 5}$
\\
$^{1}$ Department of physics, University of Warwick, Gibbet Hill, Coventry CV4~7AL, UK\\
$^{2}$ Instituto de F\'isica y Astronom\'ia, Universidad de Valpara\'iso, Av. Gran Breta\~na 1111, Valpara\'iso, Chile\\
$^{3}$ Departamento de F\'isica, Universidad T\'ecnica Federico Santa Mar\'ia, Avenida Espa\~na 1680, Valpara\'iso, Chile\\
$^{4}$ Millennium Nucleus for Planet Formation, NPF, Valpara{\'i}so, 2340000, Chile\\
$^{5}$ Centre for Exoplanets and Habitability, University of Warwick, Coventry CV4 7AL, UK\\
}

\date{Accepted XXX. Received YYY; in original form ZZZ}

\pubyear{2022}

\begin{document}
\label{firstpage}
\pagerange{\pageref{firstpage}--\pageref{lastpage}}
\maketitle

\begin{abstract}
The recently discovered system Gaia\,0007$-$1605 consisting of a white dwarf with a close brown dwarf companion and a distant white dwarf tertiary very much resembles the triple system containing the first 
transiting planet candidate around a white dwarf ever discovered:
WD\,1856+534. We have previously argued that the inner binary in WD\,1856+534 most likely formed through common envelope evolution but triple star dynamics represent an alternative scenario. 
Here we analyze different formation scenarios for Gaia\,0007$-$1605. We reconstructed the potential common envelope evolution of the system and find that assuming standard parameters for the energy budget provides a reasonable solution. In agreement with other close white dwarf $+$ brown dwarf binaries, and in contrast to WD\,1856+534, no energy sources other than orbital energy during common envelope evolution are required to understand the current configuration of the system. In addition, using analytical prescriptions for triple dynamics, we show that Von Zeipel--Lidov--Kozai oscillations might have trigger tidal migration due to high eccentricity incursions ($e\gtrsim0.997$). We conclude that the inner binary in Gaia\,0007$-$1605, as its sibling WD\,1856+534, formed either through common envelope evolution, triple dynamics or a combination of both mechanisms. 
\end{abstract}

\begin{keywords}
binaries: close -- white dwarfs -- brown dwarfs
\end{keywords}



\section{Introduction}

White dwarfs that are members of close binaries\footnote{through the paper we use the term ``close binary'' to refer to binaries with orbital periods up to hundred of days.} are important in a wide range of astrophysical contexts 
including studies of supernovae\,Ia or the detection of gravitational waves. 
The classical formation scenario for close binary stars containing a white dwarf is common envelope evolution \citep{Paczynski76}.

Common envelope evolution is an inherently complicated process and it has so far turned out to be impossible to cover the large range of spatial and temporal scales in hydrodynamic simulations \citep{ivanova_2013}. 
To compare observations and model predictions, proper simulations of the process are therefore often replaced by a parameterized energy equation relating the binding energy of the envelope and the change in orbital energy of the binary.

Observed samples of close white dwarf binaries with main sequence star companions can be used to constrain the energy budget during common envelope evolution. In the vast majority of cases, observed populations can be understood using a small common envelope efficiency and without assuming contributions from additional energy sources such as recombination energy. This finding holds for post common envelope binaries with M-dwarf  \citep{zorotovicetal2010}, substellar \citep{zorotovic+schreiber22}, as well as sun-like \citep{Hernandez21,Hernandez22,Hernandez2022b} companions. 

However, the general success of the common envelope scenario in explaining the observed populations of close binaries does not exclude the existence of alternative formation scenarios of close white dwarf binaries. 
Recently, \citet{lagosetal22-1} showed that the period distribution of white dwarfs with close G-type companions can be explained  
if systems with periods of months to years form through stable but non-conservative mass transfer \citep[see also][]{masudaetal19-1}. 
The situation is similar for the observed population of double white dwarfs, i.e. their characteristics can be best explained by considering the possibility of common envelope evolution and stable non-conservative mass transfer \citep[e.g.][]{nelemans00, Webbink_2008,woodsetal12-1}. 

Another alternative for forming close white dwarf binaries is triple dynamics. The inner binary of a hierarchical triple system may exchange angular momentum with the orbit of the distant tertiary through the so-called Von Zeipel-Lidov-Kozai (ZLK)
 mechanism \citep{VonZeipel1910,Lidov_1962,kozai_1962}, which may generate large eccentricities and subsequent tidal decay in the inner binary. 
ZLK oscillations have been used to explain a large variety of phenomena, including the formation of hot Jupiters 
\citep[e.g.][]{wu+murray03-1,naozetal11-1,petrovich15-1} or blue stragglers \citep[e.g.][]{perets+fabrycky09-1} and may even result in the merger of double white dwarf binaries \citep[e.g.][]{thompson11-1}. 

So far, however, the formation of close white dwarf binaries through ZLK oscillations remains hypothetical. To the best of our knowledge, a triple system where an inner binary containing a white dwarf has formed most likely through triple dynamics and tidal migration has not yet been identified. 
The perhaps most promising candidate known is the transiting gas giant planet around WD\,1856+534 which is part of a hierarchical triple system \citep{Vanderburgetal20}. While ZLK oscillations could in principle be responsible for the currently observed configuration \citep[e.g.][]{Munoz2020,Oconnor2021,stephanetal20-1}, it has also been shown that common envelope evolution can explain its tight orbit \citep{lagosetal21-1,Chamandy2021,Merlov2021}.
Measuring the mass of the transiting planet could constrain the evolutionary history of the system. Whilst common-envelope evolution is disfavoured for masses below $5$\,\Mjup\, ZLK oscillations also cover lower masses being the most likely scenario below $3$\,\Mjup.
Unfortunately, according to the lower limits derived from transmission spectroscopy that appear in the literature ($0.84$ and $2.4$\,\Mjup\, at the $2\sigma$ level, \citealt{Alonso2021,Xu2021}) both scenarios remain plausible.

Recently, a system very similar to WD\,1856+534 has been found and characterized by \citet{Rebassaetal22}. They showed that the infrared excess of the white dwarf Gaia\,0007$-$1605 is caused by a brown dwarf companion and
using spectral fits combined with {\it Gaia} photometry they
convincingly constrained most of the parameters of both components, with exception of the brown dwarf mass and the total age of the system (see Table \ref{tab:WD+BD params}). \citet{el-badryetal21-} and \citet{rebassa-mansergasetal21-1} furthermore revealed that the close white dwarf plus brown dwarf (WD+BD) binary is in fact the inner binary of a hierarchical triple system by discovering a common proper motion white dwarf companion. The distant white dwarf was characterized based on photometry only and the estimated parameters -- although in agreement with previous rough estimates \citep{gentile-fusillo21-1} --  are therefore less certain. The total age of the system derived from the parameters of the tertiary is $\sim10$\,Gyr which is in agreement with the likely membership of the system to the Galactic disc. 

We here present an investigation of the potential evolutionary history of Gaia\,0007$-$1605 using standard prescriptions for common envelope evolution and analytical approximations describing possible ZLK oscillations. We find that common envelope evolution represents a natural explanation for the current configuration and that the formation of Gaia\,0007$-$1605 can be reproduced with a small common envelope efficiency, such as virtually all post common envelope binaries. Tidal migration induced by ZLK oscillations also represents a plausible scenario as long as the brown dwarf survives the post-main-sequence evolution of its host star and its eccentricity reaches values greater than $\approx 0.997$.

\begin{table}
	\caption{Stellar parameters and ages for Gaia\,0007$-$1605 reported by \citet{Rebassaetal22}. Values with ``$*$'' require further confirmation. In particular, the total age of the systems is based on the sum of the cooling age of the outer white dwarf and its main sequence progenitor lifetime, which in turn depends on the initial-to-final mass relation. However, the latter is not well defined in the mass range of the outer white dwarf. The mass of the brown dwarf is based on its L$3\pm1$ spectral type.}
	\resizebox{\columnwidth}{!}{\begin{tabular}{llllcc}
	\hline
	Parameter &  Inner WD & Brown dwarf& Outer WD \\
	\hline
	Mass [$\Msun$] & $0.54\pm0.01$ &$0.07^*$ & $0.56\pm0.05$ \\
	Orbital period [d] & $1.0446\pm0.0015$ &$1.0446\pm0.0015$&-\\
	Cooling time [Gyr] & $0.360\pm0.002$ & -& $8.2\pm0.2$\\
	Total age [Gyr] & $\simeq 10^*$ & $\simeq 10^*$ & $\simeq 10^*$\\
	Projected separation [au] & - &- &1673.11\\
	
	\hline
	\end{tabular}}
	\label{tab:WD+BD params}
\end{table}

\section{Reconstructing the evolution of the inner binary with common envelope evolution}
\label{section:CEE}
We reconstructed the evolutionary history of the inner WD+BD binary in Gaia\,0007$-$1605 assuming that it evolved to its current short orbital period during a common-envelope phase and without influence of the distant companion on the orbital period decrease. We used the same method recently described in \citet{zorotovic+schreiber22} for a sample of well characterized close WD+BD binaries. In summary, we first calculated the period the system had immediately after ejecting the envelope based on the white dwarf cooling age and angular momentum loss in the post-common-envelope phase through gravitational radiation only \citep{schreiber+gaensicke03}. We then searched for possible progenitors of the white dwarf within a grid calculated with the single-star-evolution (SSE) code from \citet{hurleyetal2000} for solar metallicity, and reconstructed the common-envelope phase using Roche-geometry and the \textit{energy formalism} developed by \citet{webbink84-1}, assuming that no energy sources other than orbital and thermal energy contributed to unbinding the envelope (for more details see \citealt{zorotovic+schreiber22}). 

\begin{figure}
 \centering
  \includegraphics[width=0.40\textwidth,angle=0]{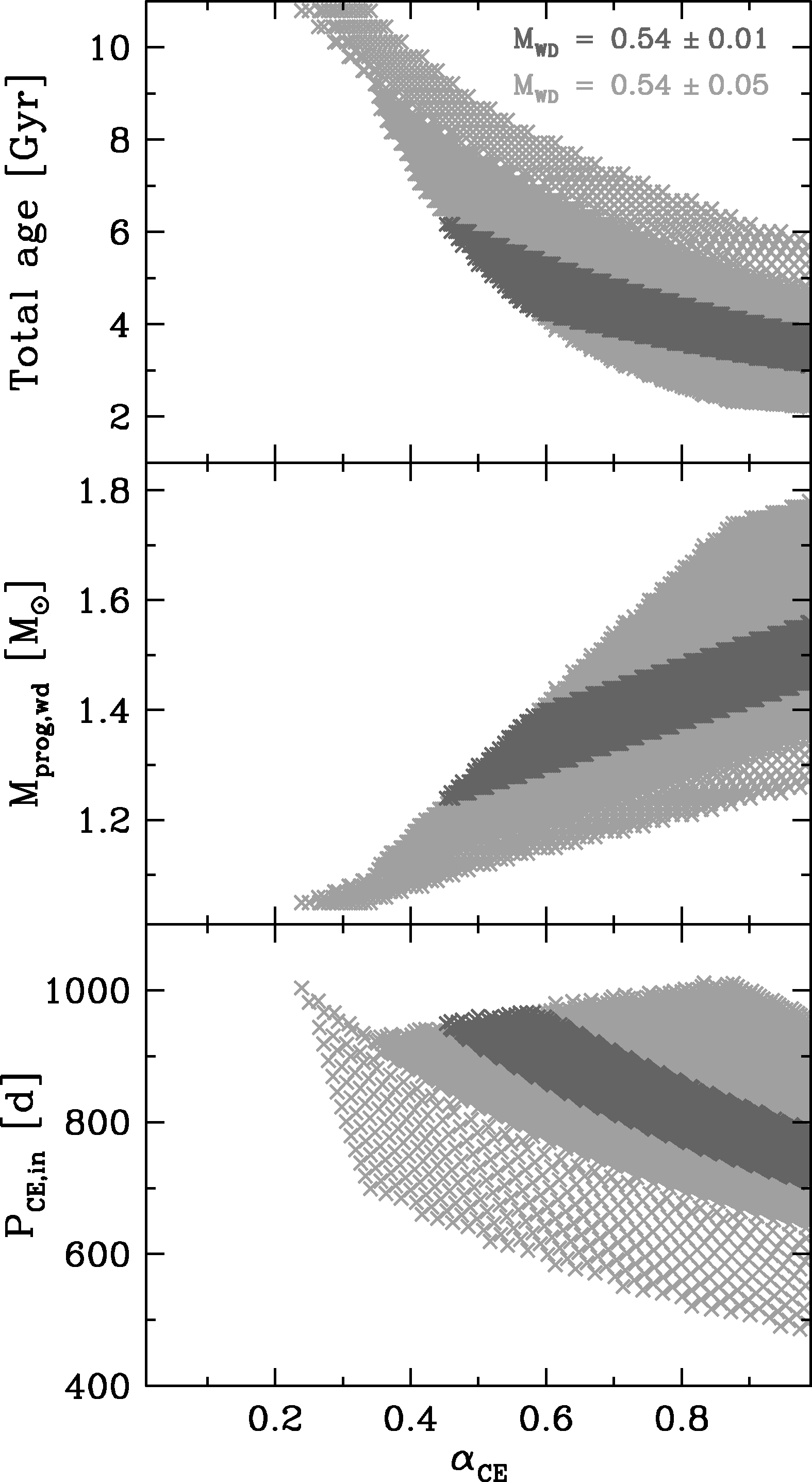}
 \caption{Total age of the system (\textit{top}), initial mass of the progenitor of the white dwarf (\textit{middle}), and orbital period at the onset of the common-envelope phase (\textit{bottom}), as a function of the common-envelope efficiency $\alpha_{\mathrm{CE}}$, for the possible progenitors of the inner white dwarf in our reconstruction.}
 \label{fig:CE}
 \end{figure}

In Fig.\,\ref{fig:CE} we present the results for the estimated total age of the system (top panel), initial mass of the progenitor of the white dwarf ($\mathrm{M}_{\mathrm{prog,wd}}$, middle panel), and orbital period at the onset of the common-envelope phase ($\mathrm{P}_{\mathrm{CE,in}}$, bottom), as a function of the common-envelope efficiency $\alpha_{\mathrm{CE}}$ (i.e. the fraction of the change in orbital energy that is used to unbind the envelope). The darker results correspond to solutions assuming a white dwarf mass within the error given by \citet[][i.e. $0.01\,\Msun$]{Rebassaetal22}, while the results marked using light gray allow the white dwarf mass to vary within a range of $\pm0.05\,\Msun$. In all cases the white dwarf progenitor filled its Roche Lobe on the asymptotic giant branch (AGB), meaning the white dwarf is composed of carbon and oxygen. Assuming that the white dwarf mass is accurate within the small error given by \citet{Rebassaetal22} we derive a maximum age of $\sim6$\,Gyr for the system, which is not consistent with the large age the authors derived based on the distant white dwarf companion. In order to obtain a larger total age of $\sim10$\,Gyr the white dwarf mass needs to be slightly smaller than estimated by \citet[][$\sim0.51$\,\Msun]{rebassa-mansergasetal21-1}, implying it descends from a low-mass progenitor ($\lappr1.1$\,\Msun) with a larger main-sequence lifetime. Also, a low-mass progenitor would have less mass in the envelope at the onset of the common envelope phase, which allows the common-envelope efficiency to be smaller, i.e. $\alpha_{\mathrm{CE}}\sim0.2-0.4$, consistent with the range of $\alpha_{\mathrm{CE}}$ derived for the sample of close WD+BD systems with accurate parameters \citep{zorotovic+schreiber22}. 

\section{Possible impact of triple dynamics on the evolution}

Given that the close inner WD+BD binary is part of a hierarchical triple, it might in principle be possible that 
ZLK oscillations had an impact on the formation of the close binary. If ZLK oscillations were present after the formation of the inner white dwarf, one cannot a priori exclude that the inner binary perhaps did not form through common envelope evolution but that instead ZLK oscillations generated large eccentricities and subsequent tidal decay, producing the short orbital period of the WD+BD inner binary we observe today. 

In what follows, we use the methodology developed by \citet{Munoz2020} to evaluate whether the formation of the close WD+BD binary can be understood by inward migration due to ZLK oscillations coupled with tidal friction after the formation of the inner white dwarf (henceforth the WD/BD + WD phase).

\subsection{Constraints for migration during the WD/BD+WD phase}
\label{section:3.1}

After the formation of the second (inner) white dwarf, migration of the brown dwarf is achieved if the inner eccentricity during ZLK oscillations is above a critical value $e_\mathrm{mig}$ required for tidal migration, but below $e_\mathrm{dis}$ to avoid tidal disruption. Both $e_\mathrm{mig}$ and $e_\mathrm{dis}$ can be obtained analytically as:

\begin{equation}
\label{eq:e_mig}
e_\mathrm{mig}=1-1.96 \left(  \frac{k_\mathrm{2,BD}\Delta t_L T}{P^2_\mathrm{BD}}\frac{M_\mathrm{WD,inner}}{M_\mathrm{BD}}\frac{R^5_\mathrm{BD}}{a^5_\mathrm{BD}} \right)^{1/7}
\end{equation}

\noindent
and 

\begin{equation}
\label{eq:e_dis}
e_\mathrm{dis}=1- \eta_\mathrm{dis} \frac{R_\mathrm{BD}}{a_\mathrm{BD}} \left( \frac{M_\mathrm{WD,inner}}{M_\mathrm{BD}} \right)^{1/3}
\end{equation}

\noindent
(Eqs.\,9 and 10 of \citealt{Munoz2020}). Here $M_\mathrm{BD}$, $R_\mathrm{BD}$, $k_\mathrm{2,BD}$, $a_\mathrm{BD}$, and $P_\mathrm{BD}$ are the mass, radius, potential Love number of degree 2, semi-major axis and orbital period of the brown dwarf, respectively. $\Delta t_L$ is the lag time, $T$ the time interval in which the migration occurs and $\eta_\mathrm{dis}=2.7$ a numerical factor to estimate the minimum orbital separation at which tidal disruption will occur \citep{Guillochon2011}.

\begin{figure}
  \begin{center}
    \includegraphics[width=\columnwidth]{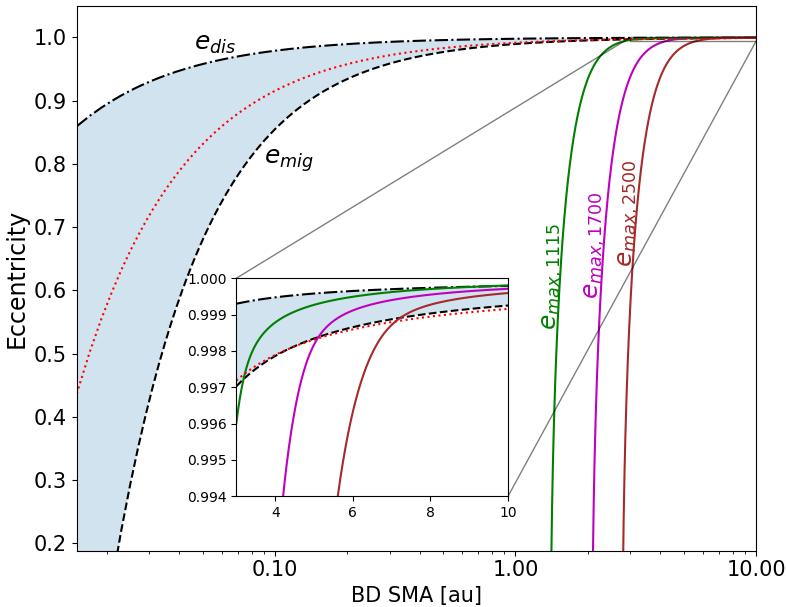}\\
  \end{center}
  \caption{Eccentricity window (filled area) in which migration would occur during the WD/BD+WD phase as function of the semi-major axis of the brown dwarf. The boundaries of this window are set by Equations \ref{eq:e_mig} (dashed line) and \ref{eq:e_dis} (dash-dotted line). The maximum eccentricity attained by the brown dwarf under the perpendicular TPQ approximation (i.e. solution of Equation \ref{eq:e_lim}) as function of the brown dwarf semi-major axis and taking $a_\mathrm{WD,outer}=1115,1700$ and $2500$\,au (green, magenta and brown lines respectively). The eccentricity required for the brown dwarf to reach its current orbital period (semi-major axis) after successful migration is given by the red dotted line which has been calculated using Equation \ref{eq:e_BD_mig}.}
\label{fig:max_e}
\end{figure}

For the inner white dwarf we assume a progenitor of $1.07\,\Msun$, which ends up as a white dwarf with $M_\mathrm{WD,inner}\approx0.52\,\Msun$ after $\approx9.7$\,Gyr according to the SSE stellar evolution code \citep{hurleyetal2000}. In the same way, we assume for the outer white dwarf a progenitor of $1.7\,\Msun$ which formed a $0.6\,\Msun$ white dwarf after $\approx2.1$\,Gyr. For the brown dwarf mass we use the value estimated by \citet{Rebassaetal22} of $0.07\,\Msun$ while its radius is assumed to be $0.086\,\Rsun$ based on the brown dwarf isochrones\footnote{\url{http://perso.ens-lyon.fr/isabelle.baraffe/COND03_models}} of \citet{Baraffe2003}. $k_\mathrm{2,BD}$ is set to $0.286$ \citep{Heller2010}, $\Delta t_L=1$\,s (i.e. 10 times the Jupiter lag-time)  and $T=0.3$\,Gyr assuming a total age of $10$\,Gyr. Figure \ref{fig:max_e} shows the range of eccentricities $e_\mathrm{mig}<e<e_\mathrm{dis}$ that allow for migration as function of $a_\mathrm{BD}$. The migration windows becomes very narrow at $a_\mathrm{BD}\simeq1$\,au, ranging from $0.988\lesssim e\lesssim0.997$.

During ZLK oscillations, extremely high eccentricities ($e\approx1$) can be achieved at the octupolar level of approximation. The importance of this regime is usually measured by 

\begin{equation}
\label{eq:af_a0_relation}
\epsilon_\mathrm{oct}=\left(\frac{M_\mathrm{WD,inner}-M_\mathrm{BD}}{M_\mathrm{WD,inner}+M_\mathrm{BD}}\right) \left( \frac{a_\mathrm{BD}}{a_\mathrm{outer}} \right) \frac{e_\mathrm{outer}}{1-e_\mathrm{outer}^{2}},
\end{equation}

\noindent where $a_\mathrm{outer}$ and $e_\mathrm{outer}$ are the semi-major axis and eccentricity of the tertiary companion. In general, it is assumed that the octupole regime plays an important role in the evolution of the system when $\epsilon_\mathrm{oct}\gtrsim0.001$. Taking a conservative value of $1700$\,au for $a_\mathrm{outer}$ (based on the projected separation of $\approx1673$\,au between both white dwarfs) and $e_\mathrm{outer}=0.5$, the octupole regime in Gaia\,0007$-$1605  becomes important when $a_\mathrm{BD}\gtrsim3$\,au. 

To estimate the maximum eccentricity $e_\mathrm{max}$ attained by the inner binary in the octuplar regime we use equation (7) of \citet{Munoz2020}, which is derived from the perpendicular test particle quadrupole (TPQ) approximation \citep[][]{Liu2015,Naoz2016} and assuming an initial circular inner orbit:

\begin{equation}
\begin{split}
\label{eq:e_lim}
0=\frac{9}{8} e^2_\mathrm{max} -\xi_\mathrm{GR} \left( \frac{1}{(1-e^2_\mathrm{max})^{1/2}}-1 \right) \\ - \frac{\xi_\mathrm{tide}}{15}\left(  \frac{1+3e^2_\mathrm{max} + \frac{3}{8}e^4_\mathrm{max} }{(1-e^2_\mathrm{max})^{9/2}} -1\right) ,
\end{split}
\end{equation}

\noindent
where

\begin{equation}
\label{eq:xi_GR}
\xi_\mathrm{GR}=\frac{3GM^2_\mathrm{WD,inner}a^3_\mathrm{outer}(1-e^2_\mathrm{outer})^{3/2}}{a^4_\mathrm{BD}c^2M_\mathrm{WD,outer}}
\end{equation}

\noindent
and

\begin{equation}
\label{eq:xi_tides}
\xi_\mathrm{tide}=\frac{15M^2_\mathrm{WD,inner}a^3_\mathrm{outer}(1-e^2_\mathrm{outer})^{3/2}k_\mathrm{2,BD}}{a^8_\mathrm{BD}M_\mathrm{BD}M_\mathrm{WD,outer}}
\end{equation}

\noindent
are terms that represent the strength of general relativistic (GR) precession and tides relative to the quadrupolar potential of the triple system, with $c$ being the speed of light. Solutions of Equation\,\ref{eq:e_lim} for the current configuration of the system are shown in Figure \ref{fig:max_e}. The value of $e_\mathrm{max}$ (magenta line) reaches the migration window when $a_\mathrm{BD}\gtrsim5.11$\,au. The eccentricities for migration correspond to a small range of rather extreme values ($e_{mig}=0.9984\lesssim e_{max}\lesssim e_{dis}=0.9998$).

The range of mutual inclinations [$90\degree-\Delta i$, $90\degree+\Delta i$] in which $e_\mathrm{max}$ is attained can be  estimated by 

\begin{equation}
\label{eq:inc_range}
\Delta i=2.9\degree \left( \frac{\epsilon_\mathrm{oct}}{10^{-3}}  \right)^{1/2}
\end{equation}

\noindent
\citep[equation 6 of ][]{Munoz2020}. For a brown dwarf located at $5.11$\,au prior to migration, $\Delta i\approx 3.6\degree$.  If the mutual inclination is taken from an isotropic distribution (i.e. uniform in $\cos{i}$), then the probability of Gaia\,0007$-$1605 being in the inclination window required for migration is $\approx6\%$. It is worth mentioning that the value of $\Delta i$ can be increased by assuming a larger inner semi-major axis but at the expense of reducing the eccentricity window for migration (unless the planet is able to survive the migration when $e_\mathrm{max}>e_\mathrm{dis}$, being partially stripped).

If migration due to ZLK oscillations is successful, the final semi-major axis of the brown dwarf can be approximated as:

\begin{equation}
\label{eq:e_BD_mig}
a_\mathrm{BD,f}\approx 2a_\mathrm{BD,0}(1-e_\mathrm{max}),
\end{equation}
\noindent
where subscripts f and 0 stand for after and before migration, respectively. We found that when Equation \ref{eq:e_BD_mig} is evaluated for solutions of equation \ref{eq:e_lim} located in the migration window, the resulting range of final semi-major axis ($0.0043\lesssim a_\mathrm{BD,f}\lesssim 0.0166$\,au) is $\approx 1\sigma$
below the one derived using $M_\mathrm{WD,inner}=0.52 \pm 0.02$, $M_\mathrm{BD}=0.07 \pm 0.01$ and the orbital separation calculated from the orbital period reported by \citet[][$0.0169\pm2.8\times10^{-4}$\,au]{Rebassaetal22}. This result is depicted in the zoomed portion of Figure \ref{fig:max_e}, in which the eccentricity required (according to equation \ref{eq:e_BD_mig}, red dotted line) for the brown dwarf to reach the observed orbital period is always below the migration window. This rather small discrepancy can be solved either by assuming a slightly lower total mass for the white dwarf/brown dwarf binary (especially a smaller brown dwarf mass than what we assumed can not be excluded given the weak current observational constraints) or by assuming in our model a smaller outer semi-major axis so the maximum inner eccentricity reaches the migration window at smaller inner semi-major axes where solutions of Equation \ref{eq:e_BD_mig} also are in this window. An example of the latter is depicted by the green line in Figure \ref{fig:max_e} which shows $e_\mathrm{max}$ when $a_{\mathrm{outer}}=1115$\,au (this value is obtained if we assume $e_\mathrm{outer}=0.5$ and the observed projected separation being the outer apoapsis distance). In this configuration, $e_\mathrm{max}$ reaches the migration window when $a_\mathrm{BD}\gtrsim3.2$\,au and the minimum eccentricity for migration is $\approx0.997$ with a similar range of mutual inclinations.

The results presented above suggest that ZLK oscillations are able to induce inward migration although two main conditions are required to make this possible. One needs (1) the inner eccentricity to be at least $\approx0.997$ which in turn implies that (2) the mutual orbital inclination should be between $\approx86\degree$ and $\approx94\degree$.  
  
\subsection{Validity of the analytical model used}

The analytical model used by \citet{Munoz2020}, developed to find the parameter space in which the tight orbit of the planet candidate in WD\,1856 can be explained by tidal migration due to ZLK oscillations, concluded that such configuration is only reproduced if the initial (i.e. during the main-sequence stage of the host and tertiary) semi-major axis of the planet is located within a narrow range of values. This result, however, is in disagreement with the outcome of the numerical simulations performed by \citet{stephanetal20-1}, who derived a much wider range of initial semi-major axis. As discussed by \citet{stephanetal20-1}, this discrepancy can be explained considering two differences in the modeling: (1) The analytical condition imposed by \citet{Munoz2020} for the planet to survive the evolution of its host are much more restrictive than the numerical simulations performed by \citet{stephanetal20-1}, (2) unlike \citet{Munoz2020}, \citet{stephanetal20-1} consider the scenario in which the planet crosses the Roche limit (i.e. the planet reaches eccentricities above $e_\mathrm{dis}$) and is partially stripped by tidal forces but survives the migration process.

Because of these limitations of the analytical prescriptions from \citet{Munoz2020}, in our analysis we simply assumed that the brown dwarf survived the evolution of its host and emphasize only the minimum eccentricity required to achieve migration. 
In addition, unlike \citet{Munoz2020} we treated the outer semi-major axis as a free parameter. In this regard it is important to recall that the outer semi-major axis used in section 3.1 serves as an example and does not exclude configurations with smaller (larger) values, which will move the maximum eccentricity attained by the brown dwarf towards smaller (larger) inner semi-major axes. For instance, taking $a_{\mathrm{outer}}=2500$\,au the minimum eccentricity for migration is  $\approx0.9989$ at $a_\mathrm{BD}\approx7$\,au as shown by the brown line in Figure \ref{fig:max_e}.

This result is consistent with the findings of \citet{stephanetal20-1} on WD\,1856 which has a very similar configuration. The range of initial semi-major axes \citet{stephanetal20-1} derived for the planet orbiting WD\,1856 peaks close to 100\,au. At such distances, $\xi_\mathrm{tide}$ and $\xi_\mathrm{GR}$ become negligible and the maximum eccentricity only depends on the mutual inclination $i$ between the inner and outer orbits ($e_\mathrm{max}=(1-5\cos^2{(i)}/3)^{1/2}$). According to Equation 8, and neglecting any orbital expansion due to stellar evolution mass loss\footnote{This assumption provides a lower limit on the true eccentricity required for migration. If the orbit expands adiabatically (i.e. the orbital separation increases by a factor $M_\mathrm{tot,0}/M_\mathrm{tot,f}$, where $M_\mathrm{tot}$ is the total mass of the inner binary and subscripts f and 0 stand for after and before mass loss) the brown dwarf/planet semi-major axis would increase by a factor of $\approx1.9$ after the formation of the host white dwarf.}, a brown dwarf/planet located at 100\,au from its host star/white dwarf would require an eccentricity of $\approx 0.9999$ to migrate to 0.02\,au, which in turn would require $i\approx89.4\degree$.

\subsection{Migration scenario due to ZLK oscillations coupled with common-envelope evolution}

\begin{figure}
  \begin{center}
    \includegraphics[width=\columnwidth]{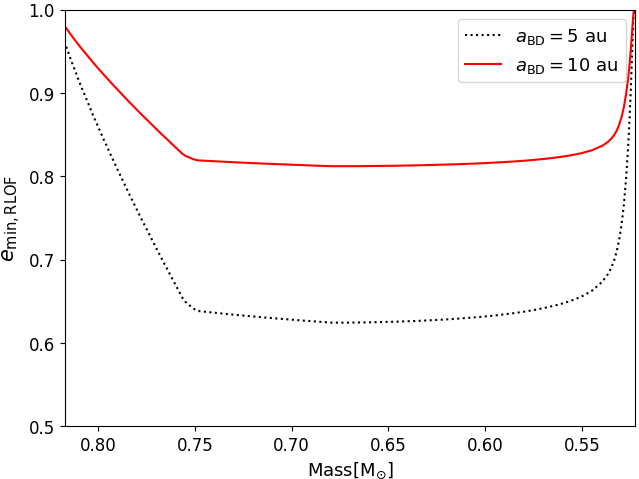}\\
  \end{center}
  \caption{Minimum eccentricity required for RLOF (Equation \ref{eq:RLOF_e}) as function of the mass of the inner white dwarf progenitor during the AGB from the SSE model. For a brown dwarf semi-major axis equal to $5$\,au ($10$\,au) the minimum eccentricity is $\approx0.62$ ($\approx0.81$) when the host star has a mass of $\approx0.67$\,\Msun\, on the thermally pulsating AGB.}
\label{fig:e_RLOF}
\end{figure}

Even if ZLK oscillations did not produce eccentricities high enough for tidal migration, lower values of the eccentricity could still have impacted the evolution and perhaps triggered common envelope evolution, similar to what has been proposed for the transiting planet around WD\,1856+534 \citep[e.g.][]{lagosetal21-1,Trani2022}. 

In order to estimate the minimum eccentricity required to start Roche lobe overflow (RLOF) we use the Roche lobe radius approximation for eccentric orbits given by \citet[][their Equation 45]{Sepinsky2007} evaluated at the periapsis of the binary and taking the Roche radius equal to the stellar radius $R_{\star}$ at the onset of mass transfer:

\begin{equation}
\label{eq:RLOF_e}
e_\mathrm{min,RLOF}=1-\frac{R_{\star}}{a_\mathrm{BD}}\frac{0.6q^{2/3} + \ln{(1+q^{1/3})} }{0.49q^{2/3}}. 
\end{equation}

\noindent
Here $q=M_\mathrm{\star}/M_\mathrm{BD}$ is the mass ratio between the host star and the brown dwarf. Both $R_{\star}$ and $M_\mathrm{\star}$ are obtained from SSE. Figure\,\ref{fig:e_RLOF} shows that for a separation of 5 (10) au, the eccentricity required to trigger mass transfer is $e_\mathrm{min,RLOF}\approx 0.62$ ($\approx0.81$) when the host star has a mass of $\approx 0.67$\,\Msun\, during the AGB phase. 

To provide one exemplary full simulation of the outlined triple star evolution, we used the \textsc{Multiple Stellar Evolution}\footnote{\url{https://github.com/hamers/mse}. MSE is based on the SSE \citep{hurleyetal2000} and binary stellar evolution \citep[BSE,][]{hurleyetal2002} codes.} \citep[MSE,][]{Hamers2021} code version 0.86, which allows to calculate the secular orbital evolution of the brown dwarf including the effects of stellar and tidal evolution, general relativity, N-body dynamics and binary interactions. Table \ref{tab:MSE_conf} summarizes the initial parameters used in our simulation. After $\approx2$\,Gyr the tertiary star evolves into a white dwarf and its orbit expands, increasing the timescale of the ZLK oscillations but keeping approximately the same maximum eccentricity. When the host star evolves through the AGB, the inner eccentricity is still high enough ($e\approx0.73$) to trigger RLOF. Although this simulation assumes an outer semi-major axis of $\approx 700$\,au (which according to the adiabatic mass loss model will increase to the observed projected separation of $\sim1700$\,au ), we verified with additional simulations that for an initial outer semi-major axis up to $1200$\,au ZLK oscillations with tidal friction are still able to lead to RLOF during the AGB.
It therefore appears to be possible that the ZLK mechanism played a role in the evolution of the triple system prior to common envelope evolution. 

\begin{figure*}
  \begin{center}
     \includegraphics[width=\textwidth]{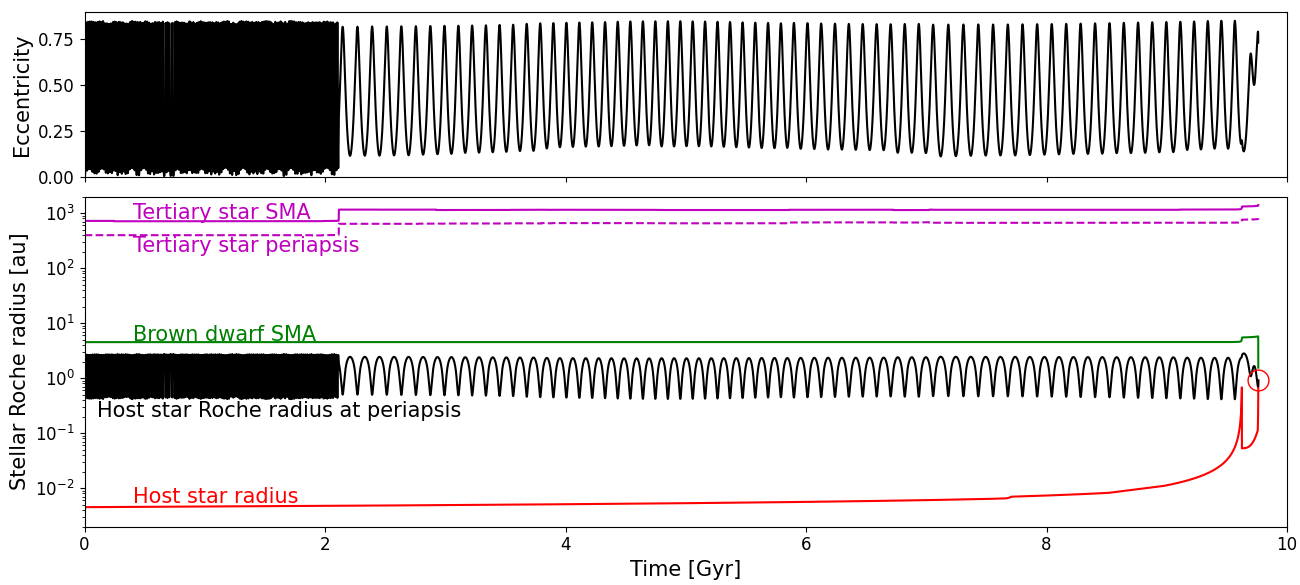}\\
  \end{center}
  \caption{Orbital evolution of Gaia\,0007$-$1605 according to the initial orbital configuration given in Table \ref{tab:MSE_conf}. The simulation is stopped when RLOF starts in the inner binary (red circle). \textbf{Top panel}: Evolution of the brown dwarf eccentricity. While the host and tertiary stars are on the main sequence ($\lesssim2$\,Gyr) ZLK oscillations produce a maximum inner eccentricity of $\approx0.85$. After $\approx2$\,Gyr the tertiary star evolves into a white dwarf, its semi-major axis increases and therefore also the timescale of the ZLK oscillations. \textbf{Bottom panel}: Evolution of the tertiary star semi-major axis and periapsis (solid and dashed magenta lines respectively), brown dwarf semi-major axis (green line), stellar radius and Roche radius at periapsis of the host star (red and black lines respectively). At $\approx9.7$ Gyr the eccentricity of the inner binary is high enough ($\approx 0.73$) to trigger RLOF (denoted by the red circle) when the semi-major axis of the brown dwarf and the tertiary star are $\approx 1.5$\,au (orbital period of $\approx 720$\,d) and $\approx1400$\,au respectively.}
\label{fig:e_oscillations}
\end{figure*}

\begin{table}
	\caption{Configuration assumed for Gaia\,0007$-$1605 at the beginning of the MSE simulation. Stellar radii and apsidal motion constants ($k_\mathrm{AM}$) are set internally by the code. For the brown dwarf we use  $R_\mathrm{BD}=0.086$\,\Rsun (expressed in au) and $k_\mathrm{AM}=0.143$ (i.e. half of the Love number of degree 2). The number of output steps is set to 20\,000. The mutual inclination between the inner and outer orbits is $65.8\degree$. }
	\resizebox{\columnwidth}{!}{\begin{tabular}{llccc}
	\hline
	Parameter & Host star & Tertiary star & Brown dwarf \\
	\hline
	Mass [$\Msun$] & 1.07 & 1.7 & 0.07 \\
	Radius [au] & Default & Default &$4\times10^{-3}$\\
	Metallicity &Solar& Solar& Solar\\
	Apsidal motion constant &0.19 &0.19 &0.143\\
	\hline
	Orbital parameter &Inner orbit& Outer orbit\\
	\hline
	Eccentricity &0&0.45\\
    Semi-major axis [au]& 4.5& 720 \\
    Inclination [rad] &0.001 & 1.15 \\
    Argument of pericentre [rad] &0.01&0.5\\
    Longitude of ascending node [rad] &0.5 & 0.5\\
    \hline
	\end{tabular}}
	\label{tab:MSE_conf}
\end{table}

\section{Conclusion}

We have studied in detail the evolutionary history of the WD+BD binary Gaia\,0007$-$1605 which is the inner binary of a hierarchical triple star system with the tertiary being a white dwarf that is $\approx8$\,Gyr older than the inner one. We found that assuming the planet survived the evolution of its host star into a white dwarf, ZLK oscillations alone can explain the configuration we observe today if the inner eccentricity reached values close or above $\approx0.997$, which in turn implies that the inner and outer orbits were (or are) close to being perpendicular to each other.

By reconstructing the close orbit of the brown dwarf we observe today through common envelope evolution we found that no energy in addition to orbital energy is required to understand the currently observed period. Our findings further support the conclusions recently drawn by \citet{zorotovic+schreiber22} that, in contrast to previous findings \citep[e.g.][]{demarcoetal11-1}, common envelope evolution with substellar companions does not require additional energy sources to play a role. This also further illustrates that in terms of the energy budget of common envelope evolution, the transiting planet around WD\,1856+534 remains an outlier as additional energy is required to reproduce the currently observed period with common envelope evolution. The idea that an additional planet might have contributed to the ejection of the envelope \citep[e.g.][]{bear_soker2011}, as suggested first by \citet{lagosetal21-1} and later investigated in more detail by \citet{Chamandy2021}, offers an elegant solution for this particular system (although it will be impossible to verify this hypothesis).

However, when considering constraints on common envelope evolution from WD+BD binaries it is important to keep in mind that Gaia\,0007$-$1605 and WD\,1856+534 might not be post common envelope systems, i.e. both common envelope evolution and/or ZLK oscillations plus tidal migration appear as potential scenarios to have produced the currently observed configuration.

\section*{Acknowledgements}
We thank the referee for helpful comments and suggestions.
This project has received funding from the European Research Council (ERC) under the European Union’s Horizon 2020 research and innovation programme (Grant agreement No. 101020057).
MRS and MZ acknowledge support from FONDECYT (grant 1221059). 
MRS also acknowledges support by ANID, – Millennium Science Initiative Program – NCN19\_171. 
This research was supported in part by the National Science Foundation under Grant No. NSF PHY-1748958 (KITP).

\section*{Data Availability}
The data and numerical tools used in this article can be obtained upon request to the corresponding author and after agreeing to the terms of use.



\bibliographystyle{mnras}
\bibliography{G0007} 




\bsp	
\label{lastpage}
\end{document}